\documentclass[prl,superscriptaddress,a4paper,twocolumn]{revtex4-1}
\usepackage{geometry}
\usepackage[english]{babel}
\usepackage[latin1]{inputenc}
\usepackage{hyperref}
\usepackage{amsmath, amssymb, amsfonts, mathrsfs}
\usepackage{graphicx}

\graphicspath{{images/}}
\usepackage{xcolor}
\usepackage{exscale}
\usepackage{graphicx}
\usepackage{amsmath}
\usepackage{latexsym}
\usepackage{amsfonts}
\usepackage{amssymb}
\usepackage{bbm}
\usepackage{xspace}
\usepackage{tabularx}

\def\pmx{\begin{pmatrix}}
\def\emx{\end{pmatrix}}

\newcommand{\ket}[1]{|#1\rangle}

\newcommand{\melvin}{{\small M}{\scriptsize ELVIN}\xspace}

\begin{document} 

\title{Generation of the Complete Four-dimensional Bell Basis}
\author{Feiran Wang}
\affiliation{Vienna Center for Quantum Science \& Technology (VCQ), Faculty of Physics, University of Vienna, Boltzmanngasse 5, 1090 Vienna, Austria.}
\affiliation{Institute for Quantum Optics and Quantum Information (IQOQI), Austrian Academy of Sciences, Boltzmanngasse 3, 1090 Vienna, Austria.}
\affiliation{Key Laboratory of Quantum Information and Quantum Optoelectronic Devices, Shaanxi Province, Xi-an Jiaotong University, Xi-an 710049, China.}
\author{Manuel Erhard}
\email{manuel.erhard@univie.ac.at}
\affiliation{Vienna Center for Quantum Science \& Technology (VCQ), Faculty of Physics, University of Vienna, Boltzmanngasse 5, 1090 Vienna, Austria.}
\affiliation{Institute for Quantum Optics and Quantum Information (IQOQI), Austrian Academy of Sciences, Boltzmanngasse 3, 1090 Vienna, Austria.}
\author{Amin Babazadeh}
\affiliation{Vienna Center for Quantum Science \& Technology (VCQ), Faculty of Physics, University of Vienna, Boltzmanngasse 5, 1090 Vienna, Austria.}
\affiliation{Institute for Quantum Optics and Quantum Information (IQOQI), Austrian Academy of Sciences, Boltzmanngasse 3, 1090 Vienna, Austria.}
\affiliation{Physics Department, Institute for Advanced Studies in Basic Sciences, Zanjan, Iran.}
\author{Mehul Malik}
\affiliation{Vienna Center for Quantum Science \& Technology (VCQ), Faculty of Physics, University of Vienna, Boltzmanngasse 5, 1090 Vienna, Austria.}
\affiliation{Institute for Quantum Optics and Quantum Information (IQOQI), Austrian Academy of Sciences, Boltzmanngasse 3, 1090 Vienna, Austria.}
\author{Mario Krenn}
\affiliation{Vienna Center for Quantum Science \& Technology (VCQ), Faculty of Physics, University of Vienna, Boltzmanngasse 5, 1090 Vienna, Austria.}
\affiliation{Institute for Quantum Optics and Quantum Information (IQOQI), Austrian Academy of Sciences, Boltzmanngasse 3, 1090 Vienna, Austria.}
\author{Anton Zeilinger}
\email{anton.zeilinger@univie.ac.at}
\affiliation{Vienna Center for Quantum Science \& Technology (VCQ), Faculty of Physics, University of Vienna, Boltzmanngasse 5, 1090 Vienna, Austria.}
\affiliation{Institute for Quantum Optics and Quantum Information (IQOQI), Austrian Academy of Sciences, Boltzmanngasse 3, 1090 Vienna, Austria.}
\email{XXX1}

\begin{abstract}
The Bell basis is a distinctive set of maximally entangled two-particle quantum states that forms the foundation for many quantum protocols such as teleportation, dense coding and entanglement swapping. While the generation, manipulation, and measurement of two-level quantum states is well understood, the same is not true in higher dimensions. Here we present the experimental generation of a complete set of Bell states in a four-dimensional Hilbert space, comprising of 16 orthogonal entangled Bell-like states encoded in the orbital angular momentum of photons. The states are created by the application of generalized high-dimensional Pauli gates on an initial entangled state. Our results pave the way for the application of high-dimensional quantum states in complex quantum protocols such as quantum dense coding.
\end{abstract}

\maketitle
Quantum entanglement is not only a curious phenomenon which radically deviates from our everyday experiences, it is also essential for many quantum information applications. High-dimensional entanglement \cite{vaziri2002experimental, dada2011experimental, agnew2011tomography, giovannini2013characterization, krenn2014generation, malik2016multi, zhang2016engineering} offers specific advantages over qubit entanglement, which is conventionally used in quantum information applications. The use of high-dimensional entanglement can enhance quantum communication schemes by increasing their channel capacity \cite{groblacher2006experimental, mafu2013higher} and offering improved robustness against sophisticated eavesdropping attacks \cite{cerf2002security, huber2013weak}. High-dimensionally entangled states can potentially be used for the teleportation of quantum states of ever-increasing complexity \cite{wang2015quantum}. In addition, such states can enhance the capacity of quantum dense-coding schemes that allow for the sharing of more information than is classically possible \cite{bennett1992communication, mattle1996dense}.

Quantum communication schemes such as these usually require control over a basis of maximally entangled quantum states---the so-called \textit{Bell basis}. For example, both quantum teleportation \cite{bennett1993teleporting} and entanglement swapping \cite{zukowski1993event} require one to unambiguously distinguish between at least one of the Bell states. In quantum dense-coding, the generation, manipulation, and discrimination of Bell states is essential, as they form the basis in which information is encoded \cite{mattle1996dense}. Polarization-entangled states have proved to be the most versatile implementation of two-dimensional Bell states thus far, and have found use is several quantum communication protocols. In addition to the polarization degree-of-freedom, two-dimensional Bell states have also been demonstrated with time-bin entangled states \cite{Brendel:1999cv}, and recently, with photons entangled in their orbital angular momentum \cite{leach2009violation, Agnew:2013ts}.

Here we demonstrate the experimental generation of a complete basis of four-dimensional Bell-like entangled states. Our experiment constitutes the first demonstration of a complete Bell basis beyond qubits. In the case of two-dimensional polarization entanglement, it is well known that one can rotate between all four Bell states with an half-wave plate (which performs a Pauli-X transformation) and quarter-wave plate (performing a Pauli-Z transformation). We generate all sixteen Bell states by generalizing this method to a higher-dimensional space.

Our states consist of two photons entangled in their orbital angular momentum (OAM) \cite{allen1992orbital, krenn2017orbital}. The sixteen orthogonal states in this basis are created by applying high-dimensional generalizations of Pauli gates on an initial OAM-entangled state that is produced via spontaneous parametric down-conversion. A four-dimensional X-gate is applied to one photon of the entangled pair, which corresponds to a cyclic transformation between the four basis states \cite{krenn2016automated,schlederer2016cyclic,babazadeh2017cyclic}. The second photon sees a four-dimensional Z-gate, which imparts a mode-dependent phase shift on the photon. We quantify the quality of our generated states by measuring their overlap with ideal states from a four-dimensional Bell basis, and verify the presence of four-dimensional entanglement by measuring an appropriate entanglement witness.
\begin{figure*}[t]
\centering
\includegraphics[width=\textwidth]{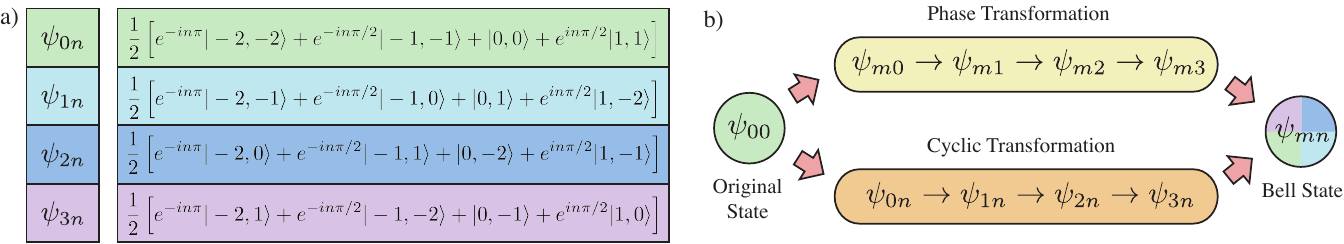}
\caption{\textbf{Generating sixteen four-dimensional Bell states.} a) The sixteen Bell states can be divided into four classes of four states each. Within each class, the states only differ in mode-dependent phases. States from different classes differ by their OAM number. b) By employing a mode-dependent phase transformation on photon A and a cyclic mode transformation on photon B, the complete set of 16 maximally entangled 4-dimensional Bell states can be obtained.}
\label{fig01}
\end{figure*}

\textit{Technique} -- The $D$-dimensional Bell basis of a bipartite system AB, as generalized in the original teleportation paper by Bennett et al \cite{bennett1993teleporting}, can be written in the form
\begin{equation}
\left| \psi  \right\rangle _{AB}^{mn} = \frac{1}{{\sqrt D }}\sum\limits_{k = 0}^{D-1} {{e^{i\frac{{2\pi }}{D}nk}}{{\left| k \right\rangle }_A}{{\left| {k \oplus m} \right\rangle }_B}} 
\label{eq01}
\end{equation}
where $ k \oplus m \equiv \left( {k + m} \right)\bmod D$. For $D=2$, this reduces to the four well-known maximally entangled quantum states ${\left| \Psi ^ \pm  \right\rangle }$ and ${\left| \Phi ^ \pm  \right\rangle }$, which are either symmetric or antisymmetric. In the four-dimensional case, equation~(\ref{eq01}) involves sixteen orthogonal Bell states that can be categorized into four distinct groups, as shown in Figure \ref{fig01}a. The four states in each group are labeled by the variable $n=0,1,2,3$, which defines the phase relationships between the probability amplitudes. As defined in Eq.~\ref{eq01}, the four-dimensional Bell basis contains two antisymmetric and six symmetric states, while the remaining eight states are neither symmetric nor antisymmetric. 

In the first set of states $\psi _{0n}$ in Figure \ref{fig01}a, photons A and B share the same state, while the relative phase between the probability amplitudes varies according to $n$. The other three sets of states are obtained by performing specific transformations on the first group. To obtain the second group $ \psi_{1n}  $, photon B is transformed using clockwise cyclic mode transformation $X$ $ \left( { - 2 \to  - 1 \to  0 \to  1 \to -2} \right)$.  For the groups $ \psi_{2n}  $ and $ \psi_{3n}  $, the state of photon B is transformed by an $X^2$ transformation $\left( -2 \leftrightarrows 0 ,-1 \leftrightarrows 1 \right) $ and an anti-clockwise cyclic transformation $X^{\dagger}$ $\left( {-2 \to 1 \to 0 \to -1 \to -2 } \right)$, respectively. To transform among states within each group, photon A undergoes a mode-dependent phase transformation. In this manner, all sixteen states in the four-dimensional Bell basis are obtained (Fig.~\ref{fig01}). Since this is a bipartite maximally entangled system, it does not matter on which photon or in which order the phase and cyclic transformations are applied.

\begin{figure*}[t]
\centering
\includegraphics[width=14cm]{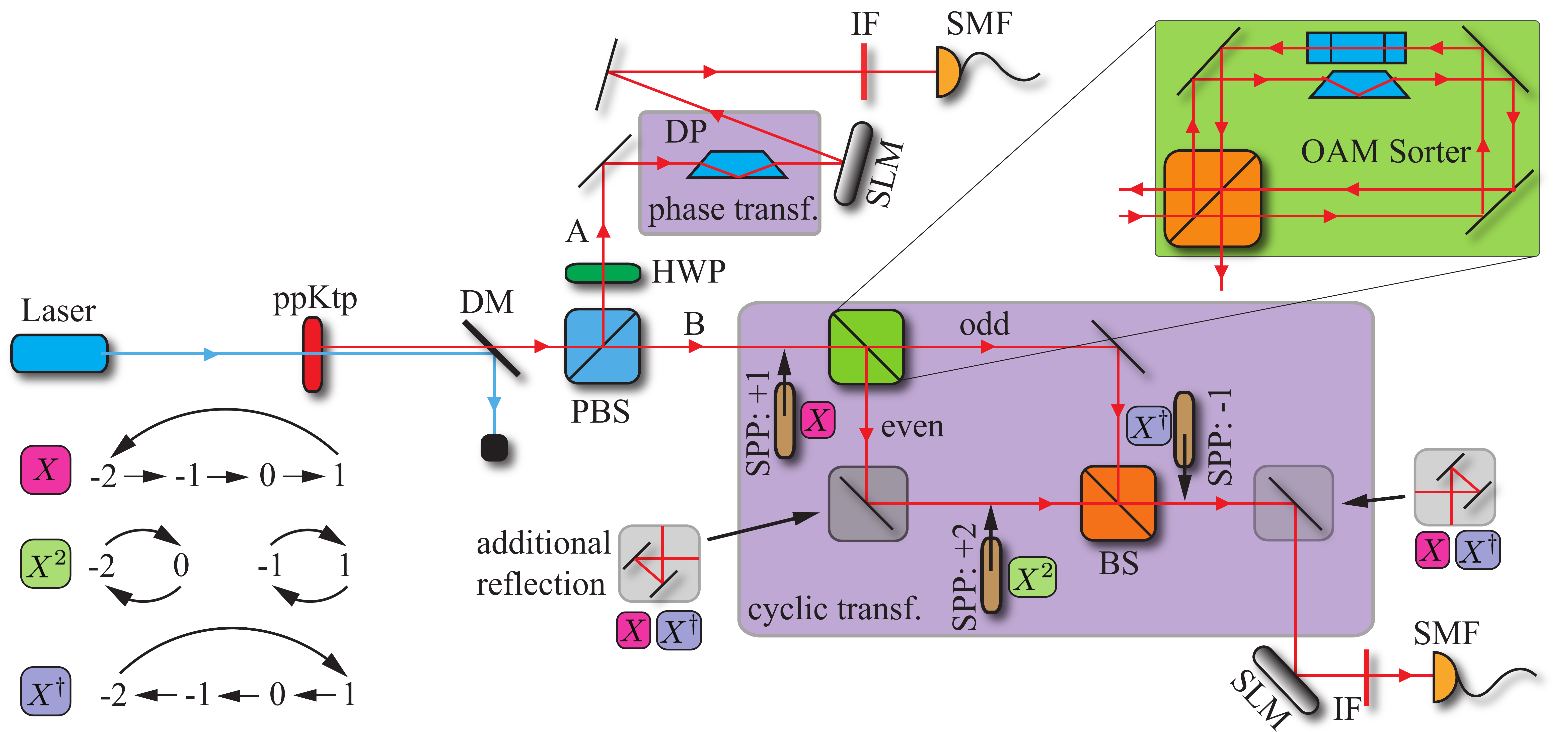}
\caption{\textbf{Experimental setup of 4d Bell states.} A laser creates a pair of OAM-entangled photons in a nonlinear crystal, and the entangled pair is deterministically separated with a polarising beam splitter (PBS). The upper path A has a Dove-prism inserted. For different rotation angles of the Dove-prism all phase transformations can be accomplished. The photons which pass through the PBS arrive at the OAM sorter (green frame). This sorter contains two Dove prisms (DP), with a relative angle of $90^\text{o}$ which results in interference that depends on the parity (odd or even) of the spatial mode. For reasons of stability, the OAM sorter is implemented as a double path Sagnac interferometer. To implement the three cyclic transformations ($X$,$X^2$,$X^{\dagger}$), only a spiral-phase-plate (SPP) and a mirror has to be installed in different positions. A spatial-light-modulator (SLM) together with a single-mode-fiber (SMF) is used to perform projective measurements.}
\label{fig02}
\end{figure*}

\textit{Experiment} -- In Fig.~\ref{fig02}, photons entangled in orbital angular momentum (OAM) are generated via a frequency-degenerate type-II spontaneous parametric down-conversion (SPDC) process in a periodically poled Potassium Titanyl Phosphate (ppKTP) crystal. The photons are produced in the state
\begin{equation}
\left| \Psi  \right\rangle {\text{ = }}\sum\limits_{\ell {=-}\infty }^{+ \infty}  {{c_\ell }} {\left| { - \ell } \right\rangle _A}{\left| \ell  \right\rangle _B},
\end{equation}
where ${\left| \ell  \right\rangle }$ represents a photon carrying an OAM of $ \ell \hbar $ and $c_\ell$ is a complex probability amplitude. For the purposes of our experiment, we use a four-dimensional subset of this state consisting of OAM mode values $\ell$ varying from $-2$ to $1$. One photon is vertically polarized and thus reflected by the polarizing beam splitter (PBS) into path A. The other photon is vertically polarized and therefore transmitted at the PBS into path A. The reflection at the PBS flips the sign of the OAM mode ($\left| \ell  \right\rangle  \to \left| { - \ell } \right\rangle $). This transforms the entangled photons into the first state of our basis $\ket{\psi_{00}}=\left( {\left| { - 2, - 2} \right\rangle  + \left| { - 1, - 1} \right\rangle  + \left| {0,0} \right\rangle  + \left| {1,1} \right\rangle } \right)/2$. In order to create the remaining three states in the first group, we apply a mode-dependent phase transformation with a Dove prism in arm A. In general, a Dove prism oriented at an angle $\alpha$ introduces a phase $\left| \ell  \right\rangle  \to \exp \left( {i2\ell \alpha } \right)\left| \ell  \right\rangle $ that depends on the OAM value $\ell$ of the incoming photon and the rotation angle $\alpha$ of the prism. The effect of this element on the state can be written as
\begin{equation}
\left| \Psi  \right\rangle  \xrightarrow{\text{$DP\left( \alpha  \right)$}} \left| {\Psi '} \right\rangle =\frac{1}{2}\sum\limits_{\ell  \in \left\{ - 2, - 1,0,1 \right\}} e^{i2l\alpha }\left| \ell  \right\rangle_A\left|  \ell  \right\rangle_B
\end{equation}
By orienting the Dove prism at different angles $(\alpha = 0,\pi /4,\pi /2, 3 \pi /4)$, we obtain all four states in one group.

A four-fold clockwise cyclic transformation of OAM modes was recently developed through the use of the computer algorithm \melvin \cite{krenn2016automated} and implemented with coherent light as well as single photons \cite{schlederer2016cyclic,babazadeh2017cyclic}. The principle idea of cyclic transformations is to split even and odd OAM modes into two different paths and manipulate them independently. Finally, the two paths are recombined coherently. In our experiment, we implement three such cyclic transformations ($X$,$X^2$,$X^{\dagger}$) at the single-photon level. 

As shown in Fig.~\ref{fig02}, we use a double-path Sagnac interferometer containing two Dove-prisms (DP)~\cite{Leach:2002wy} to split even and odd OAM modes into two different paths (green frame). In the path for even OAM modes, different OAM manipulations are performed which are necessary for the three cyclic transformations. The two paths are probabilistically recombined with a beam-splitter (BS) which forms a Mach-Zehnder (MZ) interferometer. In principle the two paths can be recombined with another parity sorter in a deterministic way. To perform the $X$ transformation $ \left( { - 2 \to  - 1 \to  0 \to  1 \to -2} \right)$ a spiral phase plate (SPP) adds an OAM quantum of $+1$ before the OAM sorter. After the sorter, one of the paths of the MZ interferometer undergoes an additional reflection. For the $X^2$ transformation $\left( -2 \leftrightarrows 0 ,-1 \leftrightarrows 1 \right) $ an SPP is inserted within the MZ interferometer adding an OAM quanta of $+2$ for even OAM modes only. An additional reflection at the end completes the $X^2$ transformation. In the case of the $X^\dagger$ transformation $\left( {  -2 \to  1 \to 0 \to -1 \to -2} \right)$, an additional reflection occurs within the MZ interferometer. Additionally, an SPP subtracting one OAM quantum the paths are recombined. Together, the three cyclic transformations on photon B and the three phase transformations on photon A allow us to obtain all four groups of states in the four-dimensional Bell basis.

\begin{figure}[ht]
\centering
\includegraphics[width=8cm]{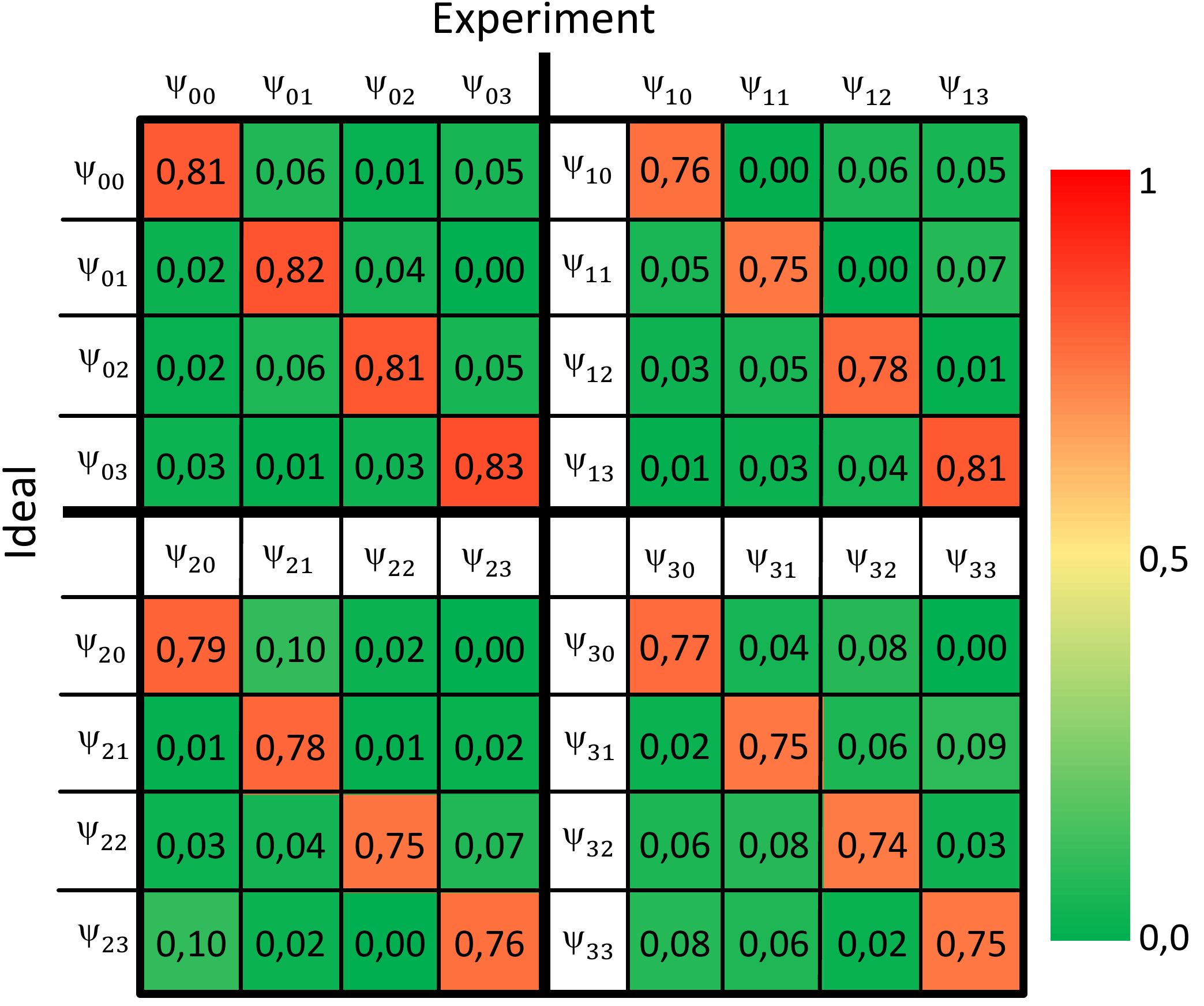}
\caption{\textbf{The overlap between generated states and ideal Bell states.} The overlap of 16 states is separated into four subgraphs as shown in the figure. The $x$ and $y$ axis in the figure represent the experimental and ideal Bell states respectively. The diagonal elements exhibit exactly the fidelity $F_{exp}$ between the experimental states and their  corresponding states.}
\label{fig03}
\end{figure}

The detection system consists of a spatial light modulator (SLM), a single-mode fiber (SMF) and a single-photon detector. The SLM is used to flatten the phase of an incoming photon, transforming it into an $\ell=0$ mode that efficiently couples to the SMF \citep{mair2001entanglement}. In this manner, the OAM content of single photons can be measured for specific modes or mode superpositions.

\textit{Results} -- The sixteen experimentally generated Bell states are analyzed using two different quantitative measures: their overlap with the theoretically expected Bell states, and a witness of four-dimensional entanglement. The overlap allows us to estimate how close we are in our experiment to the ideal Bell basis, and the witness allows us to verify the presence of genuine four-dimensional entanglement in our generated states. Fig.~\ref{fig03} shows the overlap of states within each of the four groups $\psi_{0n}$-$\psi_{3n}$. The overlap is measured by calculating the fidelity $F_{exp}=\text{Tr}(\rho_{exp} |\psi_{mn}\rangle\langle\psi_{mn}|)$, where $\rho_{exp}$ denotes the experimentally created state and $|\psi_{mn}\rangle$ the ideal Bell states. Taking the non-flat spiral bandwidth~\cite{torres2003quantum} of the SPDC state into account, the maximum expected Fidelity is limited to $93\%$. The average fidelity to the ideal state for the first group (without any cyclic transformation) is $82.1\%\pm1.1\%$. The decrease of the measured fidelity of about $11\%$ is mainly due to inter-modal cross-talk. The other three groups combined show an average fidelity to the ideal state of $76.6\%\pm2.2\%$. This shows that the cyclic transformation lowers the average fidelity by approximately $5.5\%$, which can be attributed to additional misalignments within the interferometers that comprise the $X$-gates.

Next, we certify the entanglement dimensionality of our generated states by using a bipartite entanglement witness for $d$-dimensional systems \citep{fickler2014interface, erhard2017quantum}. We search for the maximal overlap of an arbitrary 3-dimensional quantum state with a maximally entangled 4-dimensional entangled state (which is not necessarily a Bell-state but might have different phases). The theoretical maximum overlap is $\bar{F}_{max}=75\%$. If we exceed this bound in the experiment, the state is (at least) 4-dimensionally entangled.

\begin{figure}[t]
\centering
\includegraphics[width=8cm]{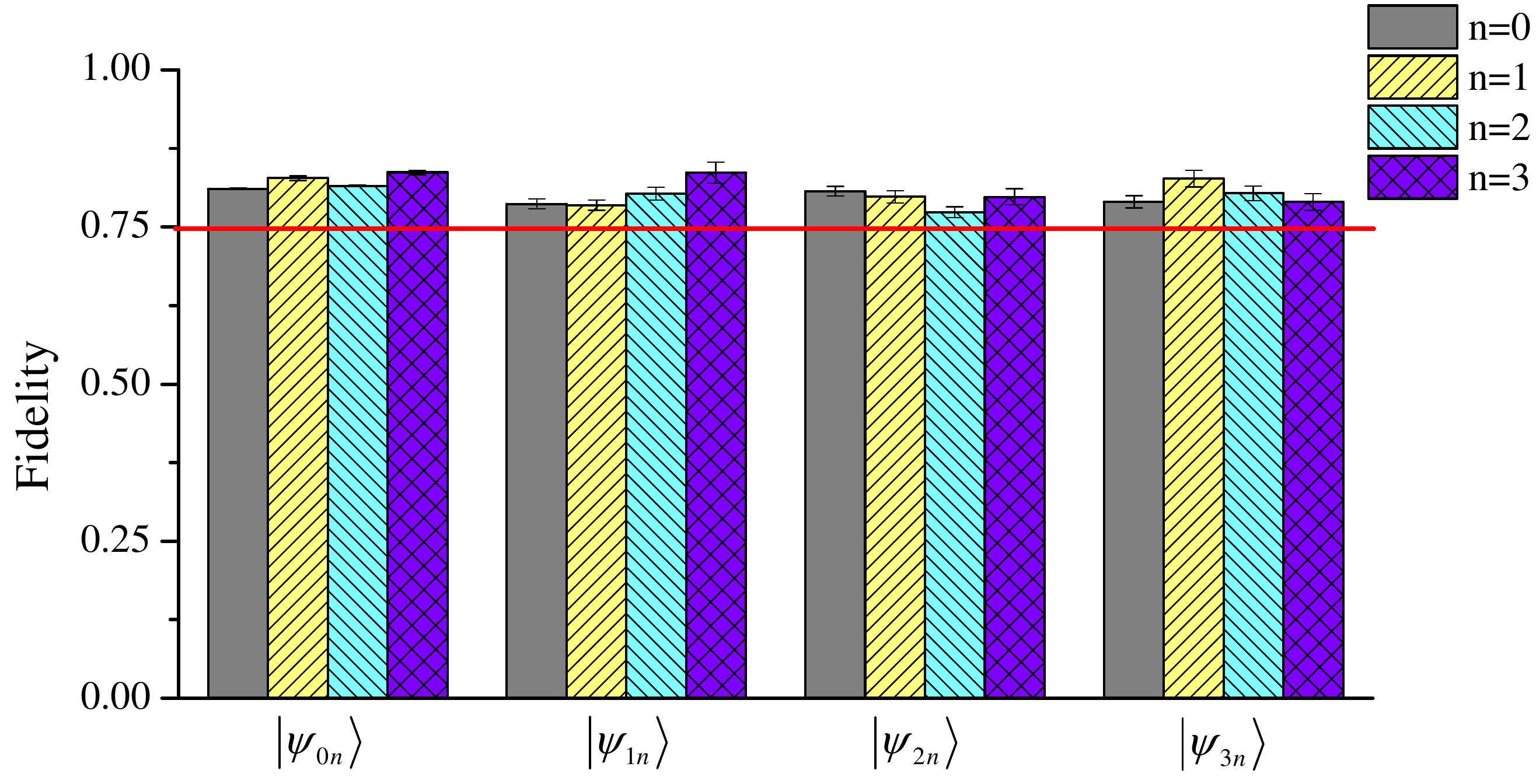}
\caption{\textbf{Fidelity Witness:} Our states exceed the bound for the overlap with a 3-dimensionally entangled state, thus it is (at least) 4-dimensionally entangled. The fidelity of 16 states is classified into four groups ($\left| {{\psi _{0n}}} \right\rangle $, $\left| {{\psi _{1n}}} \right\rangle $, $\left| {{\psi _{2n}}} \right\rangle $, $\left| {{\psi _{3n}}} \right\rangle $) in the figure. The error is calculated using Monte Carlo simulation, and the red line denotes the theoretical bound for four-dimensional witness.}
\label{fig04}
\end{figure}

The measured fidelity witnesses $F_{wit}$ for all sixteen states are plotted in Fig.~\ref{fig04}. Each of the sixteen Bell states individually exceeds the bound of 0.75 by at least three standard deviations, and is thus certified to be four-dimensionally entangled. The error in the fidelity is calculated by propagating the Poissonian error in the photon-counting rates via a Monte Carlo simulation. 

\textit{High-dimensional quantum dense coding} -- By replacing the spiral phase plates in our experiment with computer generated holograms implemented on SLMs, our technique can be extended for rapidly switching on-demand between all sixteen states in the four-dimensional Bell basis. This would constitute the first step in a high-dimensional quantum dense coding protocol \cite{mattle1996dense}: by implementing both phase and cyclic transformations on one photon, Bob can encode 4 bits of information using the two-photon four-dimensional Bell basis. The subsequent step where Alice must distinguish between all sixteen Bell states in order to decode this information provides a significant challenge. It has been shown that it is impossible to unambiguously discriminate a single high-dimensional Bell state from the others with just linear optics \cite{calsamiglia2002generalized}. However, it is possible to sort 16 Bell states into 7 classes of states that can be distinguished with a linear optical setup, as was recently demonstrated with hyperentangled time-polarization states \cite{Hill:2016jq}. Our experimental technique solely involves the photonic spatial degree of freedom, and can thus be readily combined with well-developed techniques for polarization \cite{barreiro2008beating} and time-bin encoding \cite{hill2016hyperdense}, allowing for a significant increase in Hilbert space dimensionality.

\textit{Conclusion} -- 
Here we have shown the application of recently developed high-dimensional quantum gates to photonic quantum entanglement. By doing so, we were able to create very general high-dimensional quantum states, for which no method of creation was known. The quantum states we created are a high-dimensional generalisation of the Bell-basis, arguably the most commonly used set of entangled quantum states in two dimensions. Access to the complete high-dimensional Bell basis allows for the exploration of strong non-classical correlations and their application in quantum information protocols such as quantum dense-coding. Furthermore, our technique can be used for the generation of complete sets of high-dimensional Greenberger-Horne-Zeilinger states~\cite{malik2016multi}.

\section*{Acknowledgments}
This work was supported by the Austrian Academy of Sciences (\"OAW), by the European Research Council (SIQS Grant No. 600645 EU-FP7-ICT) and the Austrian Science Fund (FWF) with SFB F40 (FOQUS) and FWF project CoQuS No. W1210-N16. F.W. was supported by the National Natural Science Foundation of China (NSFC Grant No. 11534008).

\section*{Appendix}\label{appendix}

\subsection{Deviations from the ideal states}

Deviations from the ideal states can be explained by three main effects: Non-equal distribution of modes form the original state, cross-talk between modes, and loss of coherence in the interferometers:
The spiral bandwidth of the OAM distribution is not flat, thus the state created in the down-conversion process is not maximally entangled. In our experiment we measure an initial state of $\ket{\psi}=\alpha|0,0\rangle+\beta|1,1\rangle+\beta|-1,-1\rangle+\gamma|2,2\rangle$ with $\alpha/\beta=0.69$ and $\alpha/\gamma=0.45$. Thus the maximum possible fidelity with a maximally entangled Bell state is limited by 93 percent. This inherent unbalancing of the created modes can be overcome with a procrustean filtering technique \cite{vaziri2003concentration, dada2011experimental}.
Another issue that lowers the fidelity is the cross-talk between different modes, and we find that in the computation basis, $\frac{\textnormal{cross-talk counts}}{\textnormal{all counts}}=0.11(3)$. The cross-talk limits the fidelity to 91 percent. These impurities mainly stem from misalignments of the OAM sorter and the Mach-Zehnder interferometer, which can be reduced by active stabilisation.
The coherence of the off-diagonal elements in this experiment has been measured to be $0.97(6)$ percent on average. Taking these three limiting factors into account the expected fidelity witness values are given by $F_{wit}=0.81(5)$. Hence, the observed average fidelity witness of $F_{wit}=0.808\pm 0.016$ is mainly due to unbalancing and cross-talk in the diagonal elements. 

\subsection{Overlap between states}
Here we show the data from which Figure \ref{fig03} has been created. It shows the overlap between different states from the same class, with the same OAM values but different phases.

\begin{center}
  \begin{tabular}{ | l || c | c | c | r | }
    \hline
                & $\psi_{0,0}$ & $\psi_{0,1}$ & $\psi_{0,2}$ & $\psi_{0,3}$ \\ \hline \hline
      $\psi_{0,0}$ & 0,810 &	0,063 &	0,011	& 0,048 \\ \hline
      $\psi_{0,1}$ & 0,024 &	0,823 & 0,041	& 0,002 \\ \hline
      $\psi_{0,2}$ & 0,015	& 0,060 & 0,818 & 0,049 \\ \hline
      $\psi_{0,3}$ & 0,027	& 0,006 &	0,032	& 0,835 \\
    \hline
  \end{tabular}
\end{center}

\begin{center}
  \begin{tabular}{ | l || c | c | c | r | }
    \hline
                & $\psi_{1,0}$ & $\psi_{1,1}$ & $\psi_{1,2}$ & $\psi_{1,3}$ \\ \hline \hline
      $\psi_{1,0}$ & 0,762 &	0,004	& 0,058 &	0,046 \\ \hline
      $\psi_{1,1}$ & 0,053 &	0,748	& 0,002 &	0,074 \\ \hline
      $\psi_{1,2}$ & 0,030 &	0,053	& 0,780 & 0,009 \\ \hline
      $\psi_{1,3}$ & 0,005 &	0,025	& 0,044 &	0,811 \\
    \hline
  \end{tabular}
\end{center}

\begin{center}
  \begin{tabular}{ | l || c | c | c | r | }
    \hline
                & $\psi_{2,0}$ & $\psi_{2,1}$ & $\psi_{2,2}$ & $\psi_{2,3}$ \\ \hline \hline
      $\psi_{2,0}$ & 0,788 & 0,100 & 0,020 & 0,003 \\ \hline
      $\psi_{2,1}$ & 0,010 & 0,784 & 0,008 & 0,017 \\ \hline
      $\psi_{2,2}$ & 0,027 & 0,041 & 0,751 & 0,066 \\ \hline
      $\psi_{2,3}$ & 0,098 & 0,020 & 0,001 & 0,764 \\
    \hline
  \end{tabular}
\end{center}

\begin{center}
  \begin{tabular}{ | l || c | c | c | r | }
    \hline
                & $\psi_{3,0}$ & $\psi_{3,1}$ & $\psi_{3,2}$ & $\psi_{3,3}$ \\ \hline \hline
      $\psi_{3,0}$ & 0,773 & 0,042 & 0,076 & 0,003 \\ \hline
      $\psi_{3,1}$ & 0,022 & 0,745 & 0,056 & 0,089 \\ \hline
      $\psi_{3,2}$ & 0,055 & 0,076 & 0,740 & 0,031 \\ \hline
      $\psi_{3,3}$ & 0,076 & 0,061 & 0,018 & 0,747 \\
    \hline
  \end{tabular}
\end{center}
From there, the average expected fidelity can be calculated to be $\bar{F}_{exp}=0.78\pm0.03$.
\subsection{Entanglement Witness}
First, we calculate the overlap $F_{wit}$ between our state and a $d$-dimensional maximally entangled target state. Then, we compute a $d$-dimensional entanglement bound $\mathcal{B}\left( d \right) = \sum\limits_{\ell  = 0}^{d - 1} {{\lambda _i}^{2}}$, which is the sum of the squares of all but the smallest Schmidt coefficient of the target state. If the overlap $F_{wit}$ exceeds the bound for a $d$-dimensional entangled state, then the measurement data can only be explained with a ($d+1$)-dimensionally entangled state.

\subsection{Combining the beams probabilistically}
In our experiments, we combine the two photon paths for photon B probabilistically. The beam splitter in Figure 2 is implemented via a half-wave plate at 45$^o$ in the horizontal arm (after which the polarisation is diagonal), and polarising beam splitter. In order to erase the \textit{which-path information}, we could use a polariser at 45$^o$. However, we use half-wave plate at 45$^o$ which rotates horizontal to diagonal, and vertical to anti-diagonal; and afterwards use the SLM as an effective polariser as the SLM only works with horizontally polarised light.
 

\begin{thebibliography}{10}

\bibitem{vaziri2002experimental}
A. Vaziri, G. Weihs and A. Zeilinger, Experimental two-photon,
  three-dimensional entanglement for quantum communication. \textit{Physical
  Review Letters} \textbf{89}, 240401 (2002).

\bibitem{dada2011experimental}
A.C. Dada, J. Leach, G.S. Buller, M.J. Padgett and E. Andersson, Experimental
  high-dimensional two-photon entanglement and violations of generalized Bell
  inequalities. \textit{Nature Physics} \textbf{7}, 677--680 (2011).

\bibitem{agnew2011tomography}
M. Agnew, J. Leach, M. McLaren, F.S. Roux and R.W. Boyd, Tomography of the
  quantum state of photons entangled in high dimensions. \textit{Physical
  Review A} \textbf{84}, 062101 (2011).

\bibitem{giovannini2013characterization}
D. Giovannini, J. Romero, J. Leach, A. Dudley, A. Forbes and M.J. Padgett,
  Characterization of high-dimensional entangled systems via mutually unbiased
  measurements. \textit{Physical review letters} \textbf{110}, 143601 (2013).

\bibitem{krenn2014generation}
M. Krenn, M. Huber, R. Fickler, R. Lapkiewicz, S. Ramelow and A. Zeilinger,
  Generation and confirmation of a (100$\times$ 100)-dimensional entangled
  quantum system. \textit{Proceedings of the National Academy of Sciences}
  \textbf{111}, 6243--6247 (2014).

\bibitem{malik2016multi}
M. Malik, M. Erhard, M. Huber, M. Krenn, R. Fickler and A. Zeilinger,
  Multi-photon entanglement in high dimensions. \textit{Nature Photonics}
  \textbf{10}, 248--252 (2016).

\bibitem{zhang2016engineering}
Y. Zhang, F.S. Roux, T. Konrad, M. Agnew, J. Leach and A. Forbes, Engineering
  two-photon high-dimensional states through quantum interference.
  \textit{Science advances} \textbf{2}, e1501165 (2016).

\bibitem{groblacher2006experimental}
S. Gr{\"o}blacher, T. Jennewein, A. Vaziri, G. Weihs and A. Zeilinger,
  Experimental quantum cryptography with qutrits. \textit{New Journal of
  Physics} \textbf{8}, 75 (2006).

\bibitem{mafu2013higher}
M. Mafu, A. Dudley, S. Goyal, D. Giovannini, M. McLaren, M.J. Padgett, T.
  Konrad, F. Petruccione, N. L{\"u}tkenhaus and A. Forbes, Higher-dimensional
  orbital-angular-momentum-based quantum key distribution with mutually
  unbiased bases. \textit{Physical Review A} \textbf{88}, 032305 (2013).

\bibitem{cerf2002security}
N.J. Cerf, M. Bourennane, A. Karlsson and N. Gisin, Security of quantum key
  distribution using d-level systems. \textit{Physical Review Letters}
  \textbf{88}, 127902 (2002).

\bibitem{huber2013weak}
M. Huber and M. Paw{\l}owski, Weak randomness in device-independent quantum key
  distribution and the advantage of using high-dimensional entanglement.
  \textit{Physical Review A} \textbf{88}, 032309 (2013).

\bibitem{wang2015quantum}
X.L. Wang, X.D. Cai, Z.E. Su, M.C. Chen, D. Wu, L. Li, N.L. Liu, C.Y. Lu and
  J.W. Pan, Quantum teleportation of multiple degrees of freedom of a single
  photon. \textit{Nature} \textbf{518}, 516--519 (2015).

\bibitem{bennett1992communication}
C.H. Bennett and S.J. Wiesner, Communication via one-and two-particle operators
  on Einstein-Podolsky-Rosen states. \textit{Physical review letters}
  \textbf{69}, 2881 (1992).

\bibitem{mattle1996dense}
K. Mattle, H. Weinfurter, P.G. Kwiat and A. Zeilinger, Dense coding in
  experimental quantum communication. \textit{Physical Review Letters}
  \textbf{76}, 4656 (1996).

\bibitem{bennett1993teleporting}
C.H. Bennett, G. Brassard, C. Cr{\'e}peau, R. Jozsa, A. Peres and W.K.
  Wootters, Teleporting an unknown quantum state via dual classical and
  Einstein-Podolsky-Rosen channels. \textit{Physical review letters}
  \textbf{70}, 1895 (1993).

\bibitem{zukowski1993event}
M. Zukowski, A. Zeilinger, M. Horne and A. Ekert, " Event-ready-detectors" Bell
  experiment via entanglement swapping. \textit{Physical Review Letters}
  \textbf{71}, 4287--4290 (1993).

\bibitem{Brendel:1999cv}
J. Brendel, N. Gisin, W. Tittel and H. Zbinden, {Pulsed Energy-Time Entangled
  Twin-Photon Source for Quantum Communication}. \textit{Phys. Rev. Lett.}
  \textbf{82}, 2594--2597 (1999).

\bibitem{leach2009violation}
J. Leach, B. Jack, J. Romero, M. Ritsch-Marte, R. Boyd, A. Jha, S. Barnett, S.
  Franke-Arnold and M. Padgett, Violation of a Bell inequality in
  two-dimensional orbital angular momentum state-spaces. \textit{Optics
  express} \textbf{17}, 8287--8293 (2009).

\bibitem{Agnew:2013ts}
M. Agnew, J.Z. Salvail, J. Leach and R. Boyd, {Generation of Orbital Angular
  Momentum Bell States and Their Verification via Accessible Nonlinear
  Witnesses}. \textit{Phys. Rev. Lett.} \textbf{111}, 030402 (2013).

\bibitem{allen1992orbital}
L. Allen, M.W. Beijersbergen, R. Spreeuw and J. Woerdman, Orbital angular
  momentum of light and the transformation of Laguerre-Gaussian laser modes.
  \textit{Physical Review A} \textbf{45}, 8185 (1992).

\bibitem{krenn2017orbital}
M. Krenn, M. Malik, M. Erhard and A. Zeilinger, Orbital angular momentum of
  photons and the entanglement of Laguerre--Gaussian modes. \textit{Phil.
  Trans. R. Soc. A} \textbf{375}, 20150442 (2017).

\bibitem{krenn2016automated}
M. Krenn, M. Malik, R. Fickler, R. Lapkiewicz and A. Zeilinger, Automated
  search for new quantum experiments. \textit{Physical review letters}
  \textbf{116}, 090405 (2016).

\bibitem{schlederer2016cyclic}
F. Schlederer, M. Krenn, R. Fickler, M. Malik and A. Zeilinger, Cyclic
  transformation of orbital angular momentum modes. \textit{New Journal of
  Physics} \textbf{18}, 043019 (2016).

\bibitem{babazadeh2017cyclic}
A. Babazadeh, M. Erhard, F. Wang, M. Malik, R. Nouroozi, M. Krenn and A.
  Zeilinger, High-Dimensional Single-Photon Quantum Gates: Concepts and
  Experiments. \textit{arXiv preprint arXiv:1702.07299} (2017).

\bibitem{Leach:2002wy}
J. Leach, M.J. Padgett, S.M. Barnett, S. Franke-Arnold and J. Courtial,
  {Measuring the orbital angular momentum of a single photon}. \textit{Phys.
  Rev. Lett.} \textbf{88}, 257901 (2002).

\bibitem{mair2001entanglement}
A. Mair, A. Vaziri, G. Weihs and A. Zeilinger, Entanglement of the orbital
  angular momentum states of photons. \textit{Nature} \textbf{412}, 313--316
  (2001).

\bibitem{torres2003quantum}
J. Torres, A. Alexandrescu and L. Torner, Quantum spiral bandwidth of entangled
  two-photon states. \textit{Physical Review A} \textbf{68}, 050301 (2003).

\bibitem{fickler2014interface}
R. Fickler, R. Lapkiewicz, M. Huber, M.P. Lavery, M.J. Padgett and A.
  Zeilinger, Interface between path and orbital angular momentum entanglement
  for high-dimensional photonic quantum information. \textit{Nature
  communications} \textbf{5}, (2014).

\bibitem{erhard2017quantum}
M. Erhard, M. Malik and A. Zeilinger, A quantum router for high-dimensional
  entanglement. \textit{Quantum Science and Technology} \textbf{2}, 014001
  (2017).

\bibitem{calsamiglia2002generalized}
J. Calsamiglia, Generalized measurements by linear elements. \textit{Physical
  Review A} \textbf{65}, 030301 (2002).

\bibitem{Hill:2016jq}
A. Hill, T. Graham and P. Kwiat, {Hyperdense Coding with Single Photons}.
  \textit{Frontiers in Optics} FW2B.2 (2016).

\bibitem{barreiro2008beating}
J.T. Barreiro, T.C. Wei and P.G. Kwiat, Beating the channel capacity limit for
  linear photonic superdense coding. \textit{Nature physics} \textbf{4},
  282--286 (2008).

\bibitem{hill2016hyperdense}
A. Hill, T. Graham and P. Kwiat, Hyperdense Coding with Single Photons.
  (Optical Society of America, 2016).

\bibitem{vaziri2003concentration}
A. Vaziri, J.W. Pan, T. Jennewein, G. Weihs and A. Zeilinger, Concentration of
  higher dimensional entanglement: qutrits of photon orbital angular momentum.
  \textit{Physical review letters} \textbf{91}, 227902 (2003).

\end{thebibliography}
\end{document}